\documentclass[preprint]{aastex}
\usepackage{natbib,subfig}
\usepackage{subfig}

\renewcommand {\deg}   {\mbox{$^\circ$}}

\newcommand   {\kms}   {\mbox{km\,s$^{-1}$}}
\renewcommand {\ga}    {\mbox{\rlap{\hbox{\lower5pt\hbox{$\sim$}}}\hbox{$>$}}}
\renewcommand {\la}    {\mbox{\rlap{\hbox{\lower5pt\hbox{$\sim$}}}\hbox{$<$}}}

\begin{document}
\pagenumbering{arabic} 
\def\kms {\hbox{km{\hskip0.1em}s$^{-1}$}} 

\def\msol{\hbox{$\hbox{M}_\odot$}}
\def\lsol{\hbox{$\hbox{L}_\odot$}}
\def\kms{km s$^{-1}$}
\def\Blos{B$_{\rm los}$}
\def\etal   {{\it et al.}}                     
\def\psec           {$.\negthinspace^{s}$}
\def\pasec          {$.\negthinspace^{\prime\prime}$}
\def\pdeg           {$.\kern-.25em ^{^\circ}$}
\def\degree{\ifmmode{^\circ} \else{$^\circ$}\fi}
\def\ut #1 #2 { \, \textrm{#1}^{#2}} 
\def\u #1 { \, \textrm{#1}}          
\def\nH {n_\mathrm{H}}

\def\ddeg   {\hbox{$.\!\!^\circ$}}              
\def\deg    {$^{\circ}$}                        
\def\le     {$\leq$}                            
\def\sec    {$^{\rm s}$}                        
\def\msol   {\hbox{$M_\odot$}}                  
\def\i      {\hbox{\it I}}                      
\def\v      {\hbox{\it V}}                      
\def\dasec  {\hbox{$.\!\!^{\prime\prime}$}}     
\def\asec   {$^{\prime\prime}$}                 
\def\dasec  {\hbox{$.\!\!^{\prime\prime}$}}     
\def\dsec   {\hbox{$.\!\!^{\rm s}$}}            
\def\min    {$^{\rm m}$}                        
\def\hour   {$^{\rm h}$}                        
\def\amin   {$^{\prime}$}                       
\def\lsol{\, \hbox{$\hbox{L}_\odot$}}
\def\sec    {$^{\rm s}$}                        
\def\etal   {{\it et al.}}                     

\def\xbar   {\hbox{$\overline{\rm x}$}}         

\def\la{\lower.4ex\hbox{$\;\buildrel <\over{\scriptstyle\sim}\;$}}
\def\ga{\lower.4ex\hbox{$\;\buildrel >\over{\scriptstyle\sim}\;$}}
\def\refitem{\par\noindent\hangindent\parindent}
\oddsidemargin = 0pt \topmargin = 0pt \hoffset = 0mm \voffset = -17mm
\textwidth = 160mm  \textheight = 244mm
\parindent 0pt
\parskip 5pt

\shorttitle{Sgr A*}
\shortauthors{}

\title{Occultation of the Quiescent Emission from Sgr A* by IR Flares} 
\author{F. Yusef-Zadeh\footnote{Department of Physics and Astronomy, Northwestern University, Evanston, Il. 
60208}\ ,
M. Wardle,
\footnote{Department of Physics and Astronomy, Macquarie University, Sydney NSW 2109, Australia}\ ,
H. Bushouse 
\footnote{STScI, 3700 San Martin Drive, Baltimore, MD  21218}\ ,
C.D. Dowell
\footnote{Jet Propulsion Laboratory, California Institute of
Technology, Pasadena, CA 91109}\ , 
and D. A. Roberts
\footnote{Adler Planetarium and Astronomy Museum, 1300
South Lake Shore Drive, Chicago, IL 60605}
}

\begin{abstract} 

We have investigated the nature of flare emission from Sgr A* during multi-wavelength 
observations of this source that took place in 2004, 2005 and 2006.  We 
present 
evidence for dimming of submm and radio flux during the peak of near-IR flares.  This 
suggests that the variability of Sgr A* across its wavelength spectrum is 
phenomenologically related. The model explaining this new behavior of flare activity 
could be consistent with adiabatically cooling plasma blobs that are expanding but also 
partially eclipsing the background quiescent emission from Sgr A*. When a flare is 
launched, the plasma blob is most compact and is brightest in the optically thin regime 
whereas the emission in radio/submm wavelengths has a higher opacity. 
Absorption in the 
observed light curve of Sgr A* at radio/submm flux is due to the combined effects of 
lower brightness temperature of plasma blobs with respect to the quiescent brightness 
temperature and high opacity of plasma blobs.  This implies that plasma blobs are mainly 
placed in the magnetosphere of a disk-like flow or further out in the flow. The depth of 
the absorption being larger in submm than in radio wavelengths implies that the intrinsic 
size of the quiescent emission increases with increasing wavelength which is consistent 
with previous size measurements of Sgr A*.  Lastly, we believe that occultation of the 
quiescent emission of Sgr A* at radio/submm by IR flares can be used as a powerful tool 
to identify flare activity at its earliest phase of its evolution.

\end{abstract}

\vfill\eject

\section{Introduction}

Observations of stellar orbits in the proximity of the enigmatic radio
source Sgr~A* located at the very dynamical center of our galaxy have
established that it is a 4 $\times 10^6$\msol\ black hole \citep{reid04,ghez08,gillessen09}.  The emission
from Sgr~A* consists of quasi-steady and variable components \citep[e.g.][]{hornstein07,dodds10,zadeh10}.
The variable component is now detected in almost all wavelength bands and
is used to estimate the physical quantities of the gas flow, the radiation
mechanism in different wavelength bands and the scale lengths at which
variable emission operates.  One of the puzzling aspect of the flaring
activity of Sgr A* is the physical nature of the variable emission in
different wavelengths and how its spectrum evolves with time.  Furthermore,
it is not clear whether flares are deeply embedded in the accretion disk of
Sgr A* or are physically distinct from the region where the bulk of the
emission arises \citep{zadeh06a,zadeh08,zadeh09,marrone08,eckart08,trap10}.

Flare emission at near-IR (IR) wavelengths is due to optically thin synchrotron
emission and the submm frequency is considered to be the dividing line
between optically thin emission at high frequencies (near-IR), and
optically thick emission at lower frequencies (submm/radio).  
To examine  the possibility that there is a causal association between 
radio/submm variability and IR flaring activity, 
we studied in detail the light curves of
flare emission from optically thick and thin plasma where there was
sufficient overlap between IR/X-ray and submm/radio wavelengths.  We
searched for signatures of optical depth effect against the background
quiescent emission at the earliest phase of the evolution of flares.

\section{Data Reduction and Results}

The primary purpose of our archival search was to find simultaneous
observations of Sgr A* at submm, radio and IR wavelengths.  We found
three experiments that had overlaps between these wavelengths bands.
Almost all the data presented here have already been published previously.

One multi-wavelength observing campaign was carried out on 2006 July 17
during which there was wavelength coverage between the Keck, Chandra, the
Caltech Submillimeter Observatory (CSO), the Submillimeter Array (SMA), and
the Very Large Array (VLA) of the National Radio Astronomy
Observatory\footnote{The National Radio Astronomy Observatory is a facility
of the National Science Foundation, operated under a cooperative agreement
by Associated Universities, Inc.} (NRAO) \citep{hornstein07,marrone08,zadeh08}.  The CSO data are reanalyzed with
a higher time sampling and the VLA data are recalibrated.  The light
curves presented at 7 and 13mm have selected data corresponding to {\it uv}
spacings greater than 100k$\lambda$.

The second multi-wavelength observations took place on 2005 July 31
employing the CSO, the SMA, the Keck II and Very Large Telescopes (VLT).
The time coverage on this day was excellent as there was a total of ten
continuous hours of observations with the Keck II and VLT telescopes at
IR wavelengths \citep{hornstein07,marrone08,meyer08}.  We searched for archival radio data from the VLA and found 7mm
(43GHz) and 13mm (12GHz) data taken for a temporal coverage of about four
hours on 2005, July 31.  These data were taken in the C configuration
using fast switching technique and a phase calibrators (0173+331, 17444-31166) to 
take out rapid  atmospheric fluctuations.
The light curves presented here are
self-calibrated in phase using data corresponding to {\it uv} spacings $>
50\rm k\lambda$.


Lastly, the campaign that took place on 2004 September 4 made simultaneous
observations with the Hubble Space Telescope (HST/NICMOS) and CSO. The
submm data are reanalyzed with a better time sampling \citep{zadeh06b}.  
The NICMOS data include photometry from three filters - F160W, F187N, and F190N -
which are centered at wavelengths of 1.6, 1.87, and 1.90$\mu$m. The aperture 
photometry
measurements have been corrected to an absolute flux scale, including an extinction
correction that assumes A$_K$=3.3 magnitudes for Sgr A*.


Three examples in which there is evidence for anti-correlation of optically thin and 
thick flare emission are presented.  
The first  set of light curves show radio observations of flaring activity
that took place on July 31, 2005.  \citet{meyer08} presented a
600min light curve by combining Keck II and VLT data.  Sgr A* was highly
active during this period showing several IR flares.  Figure 1a shows
IR and radio light curves.  The bottom panels show individual 7mm and
13mm light curves in two different colors.  Radio measurements indicate a
remarkable trend that the emission is higher when there is a low-level
flaring activity and is lower when there is a strong IR flare.  This
trend is best noted at times less than 5h UT and greater than 6h UT. The
flux changes by about 150 and 100 mJy at 7 and 13mm wavelengths,
respectively.  A more detailed view of flaring activity between 5.5h and
7.5h UT is shown in Figure 1b.  The light curves at at 2.2 $\mu$m, 7 and 13mm
are presented with sampling time of 60s, 30sec and 30sec, respectively.
The anti-correlation is evident  clearly, first by the rise and fall
of IR and radio activity during 5.5 and 7.5h UT. A more detailed
examination of the radio emission anti-correlating with two of the
strongest IR flares can best be viewed near 6h UT. Radio flux rises
between  the two peaks of flare emission at 5.8h and 6.2 UT. The fast
fluctuations in the flux of Sgr A* at radio wavelengths appears to track
the variation of flux at IR wavelengths. 
To clarify  this relationship better,  radio flux of 
Sgr A* are  reversed before the change in radio flux is presented in Figure 1c.
The "correlation" of radio  and IR flux is shown in remarkable 
clarity, given that the quiescent flux  and 
the net variable flux  are different at 7 and 13mm  wavelengths. 
The apparent discrepancy between 7mm and 2.2$\mu$m variation near 7.1h UT 
may be due to poor time sampling $\sim 3$min of Keck  data and a sudden change 
of observations from  VLT to Keck
which starts  around 6.8h UT. The anti-correlation between near-IR and 
1.3mm data (Marrone et al. 2008) was not as clear as that of radio and 
near-IR data but 1.3mm  
data  could suffer from poor sampling and complication from time delayed 
emission. 
To examine the anti-correlation statistically, 
Figure 2a,b show the cross correlation of the IR data with 7mm and the merged 
7 and 13mm data,  respectively. 
The  cross-correlation analysis  use the ZDC function algorithm \citep{alexander97}. A minimum in the  likelihood value is identified at a zero time lag, strengthening the 
anti-correlation  of IR and radio data.

The second  evidence for dimming of optically thick 
light comes from observations made in 2006 July 17 \citep{marrone08,zadeh08}.  The authors showed evidence of a time delay between the peaks of 
IR/X-ray emission with respect to radio/submm emission.  Figure 3a shows simultaneous 
light curves of Sgr A* at X-rays between 2 and 8 keV, IR K$'$ band \citep{hornstein07}, 850$\mu$m, 1.3mm, and 7mm.  
In these observations, as noted 
in the top panel of Figure 3a, the second half of the X-ray flare was observed with Keck 
and showed evidence of flaring at IR wavelengths. 
This figure shows a dip in the flux of mm and submm 
emission, during which an X-ray flare is detected.  
A close-up view  during the peak emission at X-rays between 5:30 and 8h UT is 
shown in Figure 3b.  The light curves at 1.3mm and 850$\mu$m track each other well.
However, we note the 7mm light curve seems to be somewhat shifted in time by with respect to 
the 1.3mm light curve. The poor sampling of 1.3mm data between 6 and 7h UT makes the comparison 
difficult. 
The largest drop in the 1.3mm flux is 
noted at 6.7h UT. The average dimming at 850$\mu$m, 1.3mm and 7mm range between 0.5-1 Jy, 
0.5 Jy and 200-300 mJy, respectively.  The 13mm data also showed a very similar light 
curve 
to that of 7mm with a flux variation of about 100-150 mJy.

Although we note a general 
trend in the anti-correlation of X-ray and optically thick emission, this 
anti-correlation should have been evident between IR and optically thick emission. 
This is because X-ray emission is processed IR emission and does not exactly tract 
the flux of a IR flare.  The duration of a typical X-ray flare is generally shorter 
than in IR wavelengths \citep{dodds09,marrone08}.  
The 2.2$\mu$m flare of Figure 3
indicates the duration of flare is longer than that observed in X-rays by at least 30 
minutes on the decaying side of the flare. Unfortunately, there is no IR data 
available at times earlier than 6:45h UT. Nevertheless, the rise of flux at 7mm, 1.3mm 
and 850$\mu$m between 6h and 6:45h UT is consistent with the decline in the IR flux. 
Using the  
anti-correlation between   IR and radio flux, we expect that 
the IR flare should 
have started around 5h UT with a flare duration of about  two hours. 
Another feature we note in Figure 3 is a possible trend that the dimming of light
appears to end first at longer wavelengths.  
However, this needs to be confirmed as SMA data are not sampled well during the dip in the quiescent 
flux of
Sgr A*.



Lastly, the light curves showing evidence of dimming at submm wavelengths is also 
detected in an observing campaign that took place on 2004, September 4.  The light 
curves presented in Figure 4a show emission at 850$\mu$m and three IR 
wavelengths at 1.6, 1.87 and 1.90$\mu$m.  Two strong IR flares and one submm 
flare are detected near 4.2h UT, 7.6h UT and 7h UT, respectively.  (see also Figure 
10 of \citet{zadeh06b}.  A previous analysis of submm data 
using long sampling time of 
20min in the construction of the light curve suggested the possibility that the submm flare is 
the counterpart either to flare at 4.3h UT delayed by $\sim$200min with respect to 
the first IR flare or to the second IR flare at 7.6 with no time delay 
\citep{zadeh06b}.  However, the new analysis shows a new feature that 
had not been recognized before.  The 850$\mu$m light curve indicates a decrease in 
the quiescent flux of Sgr A* by $\sim$ 500 mJy during which the second IR 
flare occurred. We also note a decrease in the submm flux at the start of IR observations 
when there is a detection of a flare.  Figure 4b 
shows detailed views of the light curves to demonstrate 
the brightening and dimming of IR and submm flux during flaring activity near 
7.6h UT, respectively. This figure provides another support for a general trend in 
the anti-correlation between IR flare emission and submm flux of Sgr A*.  
Exceptions to this trend are two data points at 7.8h UT during which submm flux is 
lower than the peak IR flux near 7.5h UT. Unfortunately, the data are not 
sampled well at these wavelengths. Another effect that may complicate the analysis 
of absorption in light curves at radio and submm wavelengths could be 
due to flaring events 
that  are time delayed and are superimposed on an absorbing feature. 


\section {Discussion}

The infrared wavelength band is the most important part of the spectrum of
flare emission from Sgr A*.  This is because the flare is first launched in
IR wavelengths followed by radio and submm emission being delayed with
respect to the peak in IR emission.  In this model, flaring at a given
frequency is produced through the adiabatic expansion of an initially
optically thick blob of synchrotron-emitting relativistic electrons.  The
expanding hot plasma model of Sgr~A* flares has explained the
nature of time delay of flare emission in the optically thick regime.
However, the relationship between IR and radio/submm flare emission
has not been well established.  
Recent work has suggested that 
that the optically thin component of flare emission is unrelated to the evolution
of optically thick emission and that that IR and radio/submm flares
emission operate independent of each other \citep[e.g.][]{sabha09}. The
light curves presented here demonstrate the dimming of the quiescent flux
of Sgr A* at radio/submm wavelengths while IR flares are observed.
The new measurements show that the observed flares in different frequencies
operate under the same physical mechanism.

A likely scenario for the drop in flux seen at radio and submm frequencies
during IR flaring is that the electron population responsible for the
flaring is optically thick at lower frequencies, partially obscures the
submm-radio emission from the accretion flow, and has a lower brightness
temperature.  To explore this picture we construct a simple two-component
model.  The first component is a homogeneous, foreground, synchrotron
emission region responsible for the IR flare.  We characterize this by an
$E^{-2}$ electron energy spectrum of relativistic electrons, a magnetic
field strength $B$, and radius $R$.  This component is responsible for the
IR flare and also for producing the dips in the emission from Sgr A*.  The
second, background, component is responsible for the quiescent flux from
Sgr A* in the radio and submm, is also homogeneous and is parametrized by
its flux density $S_{q\nu}$ and radius $R_{q\nu}$, both of which are
frequency-dependent.  For simplicity, we assume that the two components are
concentric, so that the fraction of the quiescent background source covered
by the flaring component is $f=(R/R_{q\nu})^2$ when $R<R_{q\nu}$ and $f=1$
otherwise.

The net flux received from Sgr A* in this simple model consists of
contributions from the flaring region, and the obscured and unobscured
fractions $f$ and $1-f$ of the quiescent source:
\begin{eqnarray}
    S_\nu &=& \Omega J_\nu\left(1-\exp(-\tau_\nu)\right) + (1-f)S_{q\nu}+ 
    f S_{q\nu}\exp(-\tau_\nu)  \\[6pt]
    &=& S_{q\nu} + (\Omega J_\nu - f S_{q\nu}) \left(1-\exp(-\tau_\nu)\right) \,,
    \label{eq:Snu}
\end{eqnarray}
where the flare region has source function $J_\nu$, optical depth
$\tau_\nu$, and subtends solid angle $\Omega = \pi R^2/d^2$ at Earth, where
$d=8\,$kpc is our adopted distance to the Galactic Center.  

The synchrotron source function for an electron number density spectrum 
$n(E)\propto E^{-2}$ 
is
\begin{equation}
    J_\nu = 0.646\; \frac{\nu^2}{c^2}\;E_\nu\,,
    \label{eq:Jnu}
\end{equation}
where 
\begin{equation}
    E_\nu = \left(\frac{4\pi m_ec\nu}{3eB}\right)^{1/2}m_ec^2
    \label{eq:Enu}
\end{equation}
is the typical energy of the electrons responsible for synchrotron emission
at frequency $\nu$, and the other symbols have their usual meaning \citep[e.g.][]{rybicki86}.
We write the optical depth as $\tau_\nu = \kappa_\nu R$, where
the synchrotron absorption coefficient is
\begin{equation}
    \kappa_\nu = \frac{2}{9} \frac{e^3 B}{m_e\nu^2}\;  n(E_\nu)\,.
    \label{eq:kappanu}
\end{equation}
Note that we have implicitly set the electron pitch angle to 90 degrees in
equations (\ref{eq:kappanu}) and (\ref{eq:Enu}).  Finally, we adopt a fixed
ratio between the energy density of the electron population, i.e.
$U_e=\int_{E_1}^{E_2}
En(E)\,dE$, and the energy density of the magnetic
field, i.e. $U_B = B^2/8\pi$, so that
\begin{equation}
    n(E) =  \frac{B^2}{8\pi \phi E^2} 
    \label{eq:equipartition}
\end{equation}
between energies $E_1$ and $E_2$, where $\phi=(U_B/U_e)\,\ln(E_2/E_1)$.  

The flux in the infrared is dominated by optically thin flare emission, 
and equation (\ref{eq:Snu})
reduces to
\begin{equation}
    S_\nu = \Omega J_\nu \tau_\nu \,,
    \label{eq:SIR}
\end{equation}
This allows us to use equations (\ref{eq:Jnu})--(\ref{eq:equipartition}) 
to compute the flare region's
magnetic field strength, optical depth and source-function flux
\begin{eqnarray}
    B &=& 38\,(\phi_3 S_{10})^{2/7} \,R_{12}^{-6/7} \quad\textrm{G}\,,
    \label{eq:Best} \\[6pt]
    \tau_\nu &=& 0.114 (\phi_3)^{-1/7} S_{10}^{\,8/7} R_{12}^{-17/7}\,
    \nu_{350}^{-3}\,,
    \label{eq:tauest}  \\[6pt]
    \Omega J_\nu &=& 1.7 (\phi_3 S_{10})^{-1/7} R_{12}^{17/7} 
    \nu_{350}^{5/2}\quad\textrm{Jy}\,,
    \label{eq:Omeganuest}
\end{eqnarray}
where we have scaled the results using $R_{12}=R/10^{12}$\,cm, $\phi_3 =\phi/3$, $S_{10} =
S_\nu/10$\,mJy at 2.2\,$\mu$m, and $\nu_{350} = \nu/350\,$GHz.  
The weak dependence on the equipartition factor $\phi$ mean that our
results are insensitive to the ratio of the energy densities in the 
magnetic field and electron population.

Having determined the properties of the flare source region, we can compute
its obscuring effect on the radio--submm quiescent emission using
(\ref{eq:Snu}).  By way of illustration we adopt $\phi = 3$, 10\,mJy flux
of at 2.2\,$\mu$m, and quiescent fluxes $S_{q\nu}=3.0$ and 1.8\,Jy at $\nu
= 350$ and 43\,GHz, respectively.  At each frequency, for a given choice of
quiescent source radius $R_{q\nu}$ we compute the flare region radius $R$
that would produce a specified drop in the total flux.  The results are
plotted in Fig.\ 5 for flare-induced drops of 0.05, 0.1, and 0.2\,Jy at 350 GHz 
(850$\mu$m) and 43 GHz (7mm).

At 350\,GHz, absorption of $\sim 0.1\,Jy$ during an IR flare implies that
both $R$ and $R_{q\nu}\approx 10^{12}$\,cm.  The latter is consistent with the
size of Sgr A* inferred from VLBI measurements at 230\,GHz, which found a major axis
FWHM $\sim 37\,\mu$as \citep{doeleman08} and in conjunction with lower
frequency measurements \citep{bower06} infer a $\nu^{1.44\pm0.07}$
wavelength dependence, implying a major axis FWHM of $\sim 20\,\mu$as at
350\,GHz, equivalent to $2.4\times10^{12}$\,cm.  While choices of smaller
$R$ in principle produce similar absorption levels, the magnetic field
strength scales as $R^{-6/7}$, and exceeds 100\,G for $R\la 3\times10^{11}
$\,cm.  

The 43\,GHz curves in  Figure  5 imply that values of $R$ much below 
$10^{12}$\,cm require  unrealistically small size for Sgr A* at 43\,GHz.
The radius of a flare $R\approx 10^{12}$\,cm , on the other hand, implies that 
$R_{q\nu}\approx
5\times10^{12} $\,cm at 43\,GHz.  While this is several times smaller than the
measured semi major axis,  $\sim 2\times 10^{13}$ cm, the semi-minor axis 
is not yet determined and may be a few times less, so this discrepancy is 
not as large as may first appear.

This simple model shows that occultation of Sgr A* by an IR flare source
can produce noticeable dips in the flux from Sgr A* at radio and sub-mm
frequencies.  The major uncertainties are the electron spectral index,
which is used to infer the optical depth at radio and submm from  the
IR flux density, and the requirement that the IR flare region must overlie
the radio emission.  Note however that for a given IR flux density, a
steeper electron spectrum implies greater optical depth at low frequencies.
This latter would more naturally occur if the flare emission occurred in a
magnetosphere of a disk-like flow or further out in the flow. 
 One would
expect that the geometry would not be favorable all of the time.

%
%
%

A recent model has  considered 
time  dependent model 
of flare emission from Sgr A* \citep{dodds10}.  
They argue  that flares 
are generated from the dissipation of the magnetic field due to the reconnection of 
the  field  in the accretion disk \citep[see also][]{yuan09}. 
This model predicts 
a dimming of flux at submm flux.  Although their picture is quite consistent with
some of the light curves presented here, more detailed observations are required to 
distinguish between these two models. In other words,  
the question is  whether the occultation  is to due to an absorbing blob of 
hot plasma against a background quiescent flux of Sgr A* or 
due to a  decrease in the flux as a 
a consequence of the dissipation of the magnetic field 
due to reconnection of magnetic  field lines.  
Future detailed modeling and 
observations with long time coverage should be able to address the difference between these 
two models. 




Acknowledgments:
This work is partially supported by the grant AST-0807400 from the NSF. 
 We are grateful to D. Marrone, S. Hornstein and 
F. Baganoff for providing us with their data.



\begin{figure}
\center
\includegraphics[scale=0.8,angle=0]{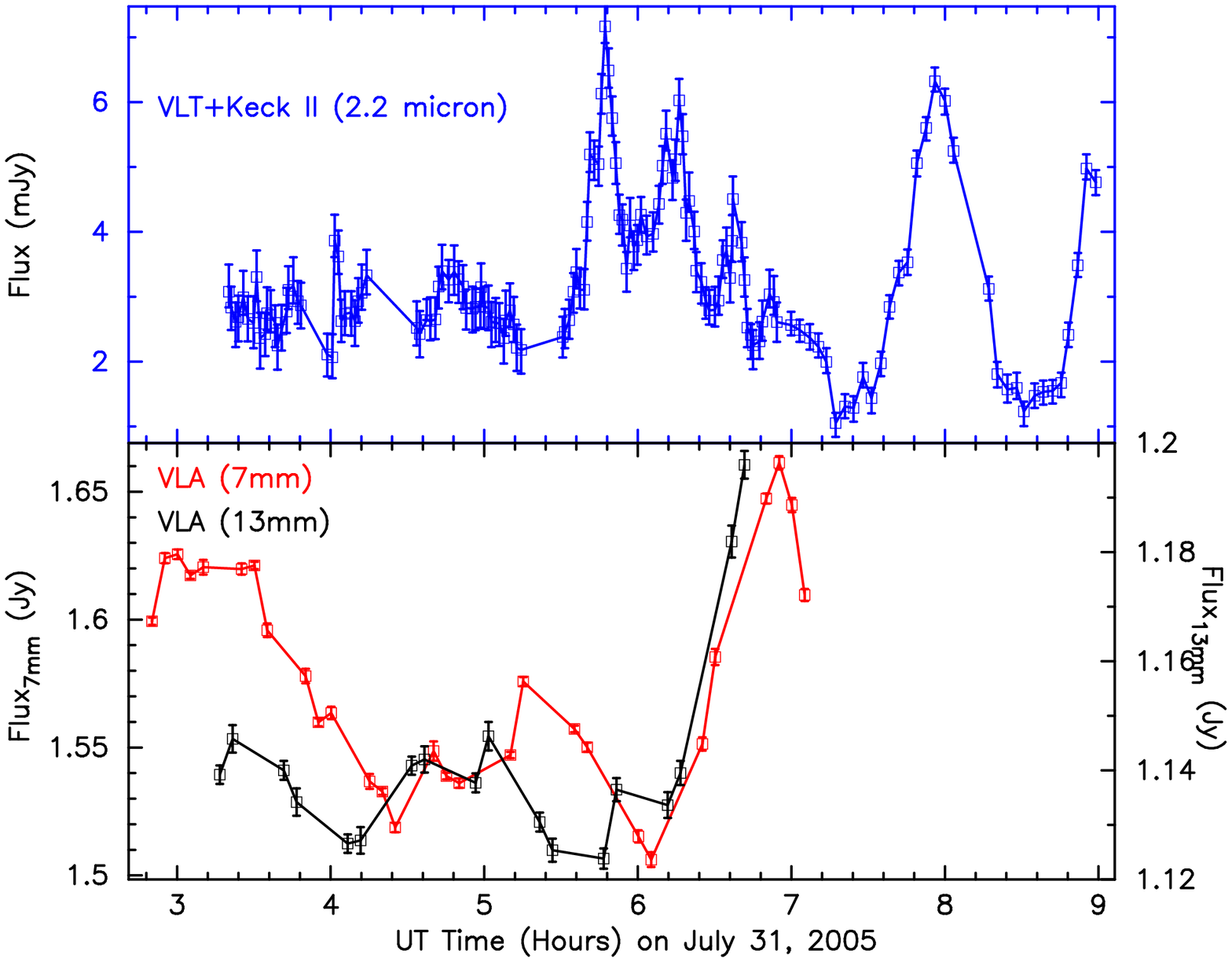}\\
\includegraphics[scale=0.4,angle=0]{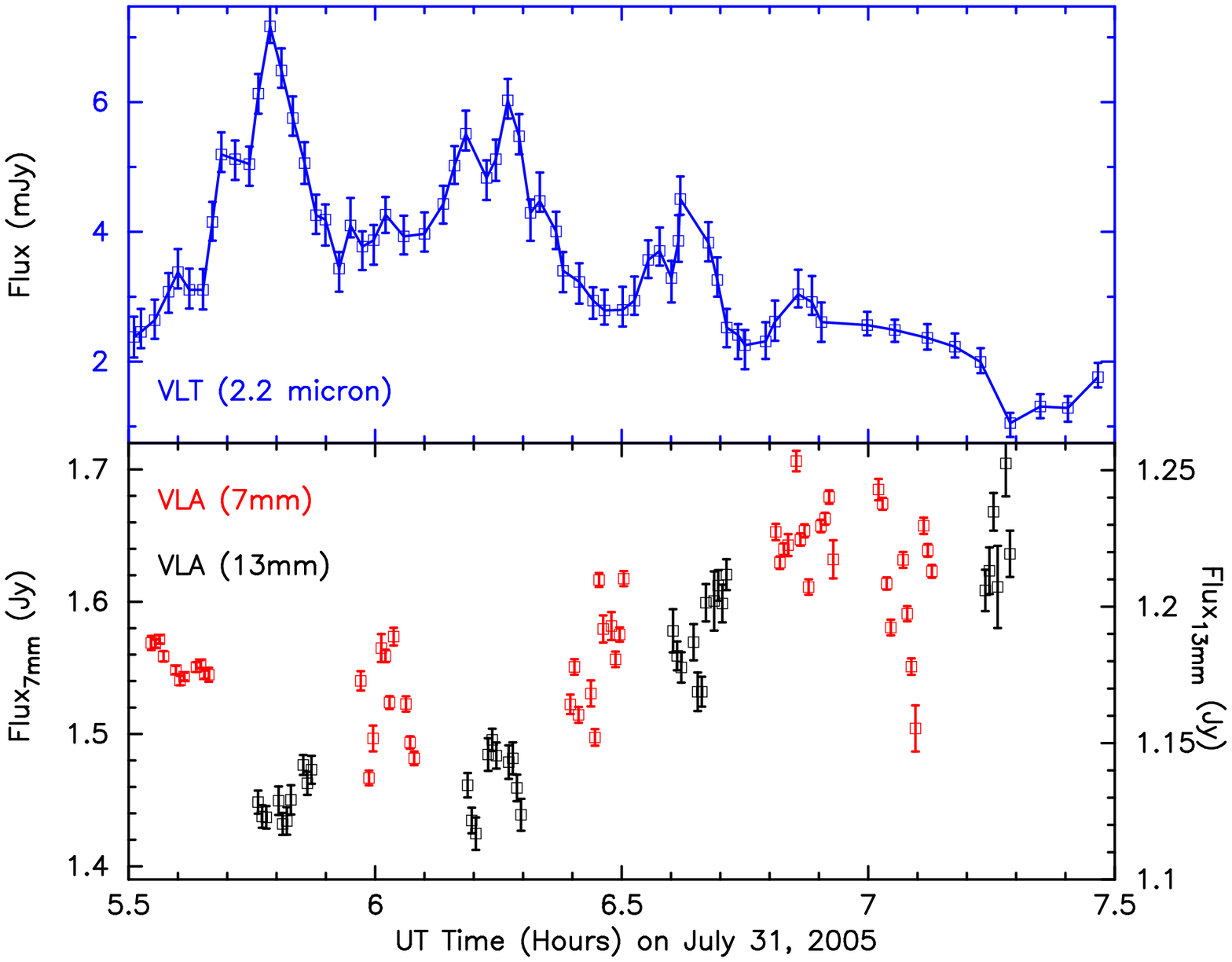}
\includegraphics[scale=0.4,angle=0]{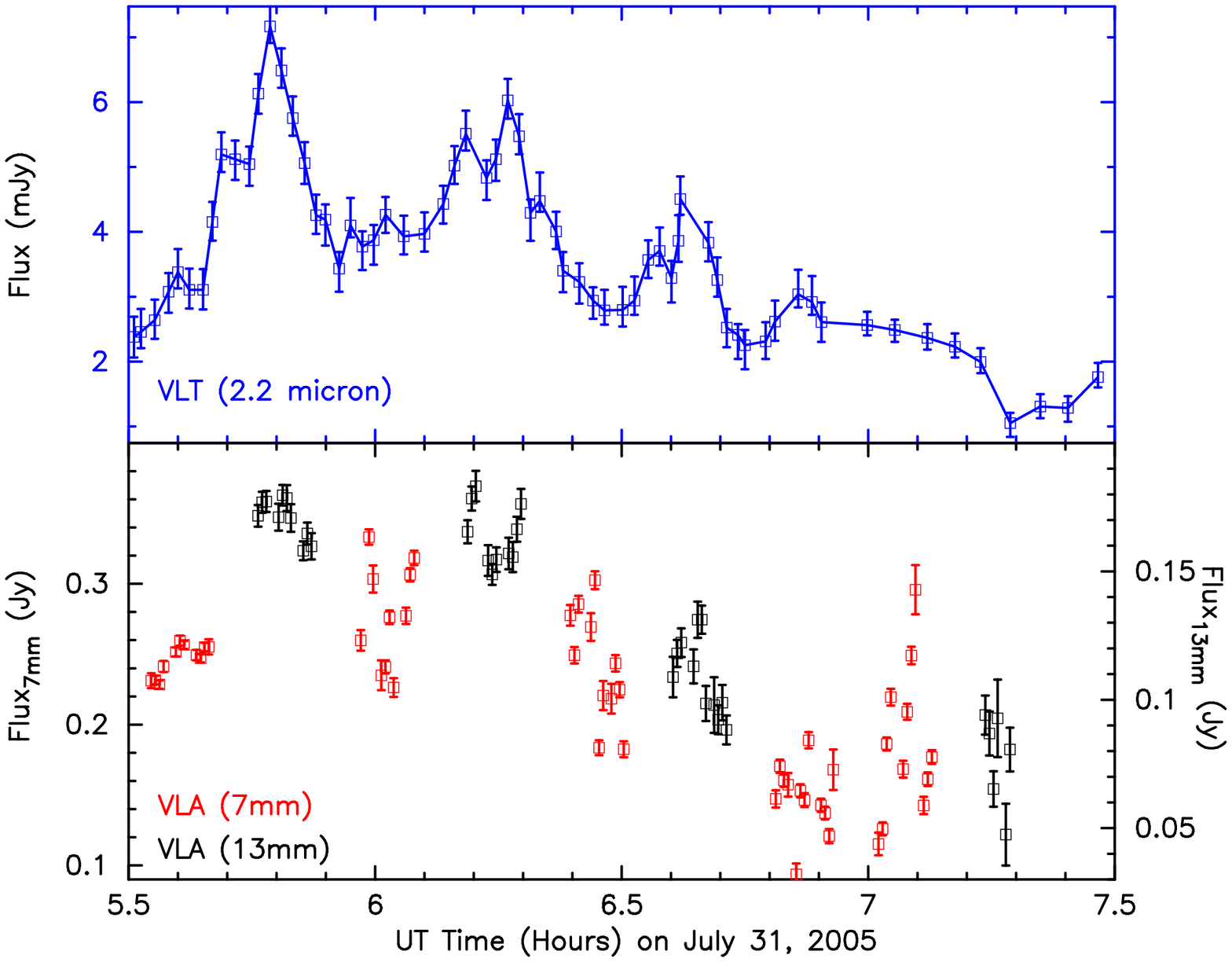}
\caption{
{\it (a) Top}
The IR data was extracted from Figure 3 of \citet{meyer08}. 
The sampling time of 7 and 13mm data is 300sec. 
{\it (b) Bottom Left}
 The same as in (a) except that 
a close-up view
is  presented.  The sampling time for the 
IR, 7mm and 13mm are about 1min, 30sec and 30sec, respectively. 
{\it (c) Bottom Right} Similar to (b) except that radio data 
are reversed by first multiplying by -1 and then adding a flux of 
1.8 and 1.3 at 7 and 13mm, respectively.  
}
\end{figure}

\begin{figure}
\center
\includegraphics[scale=0.4,angle=0]{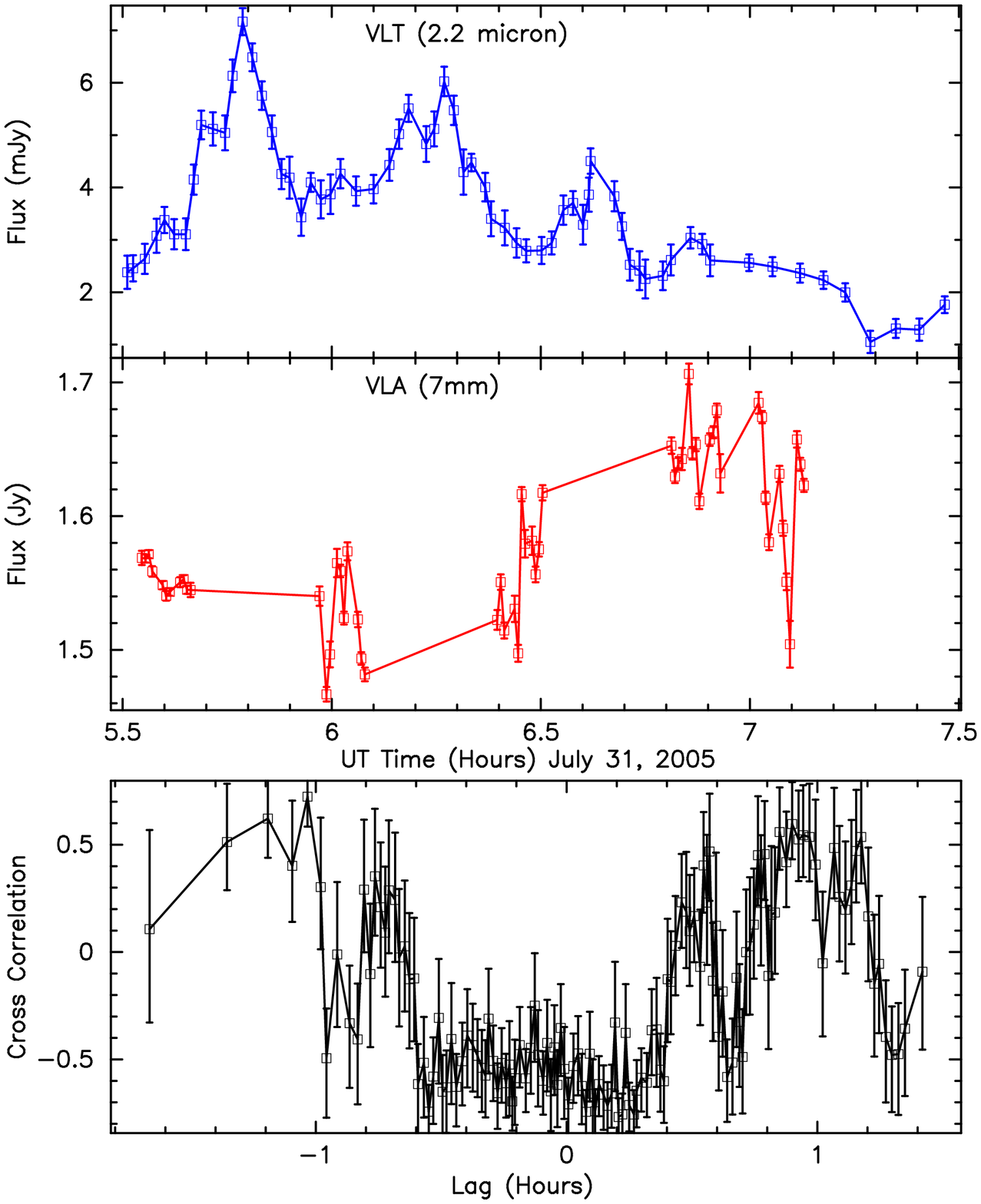}
\includegraphics[scale=0.4,angle=0]{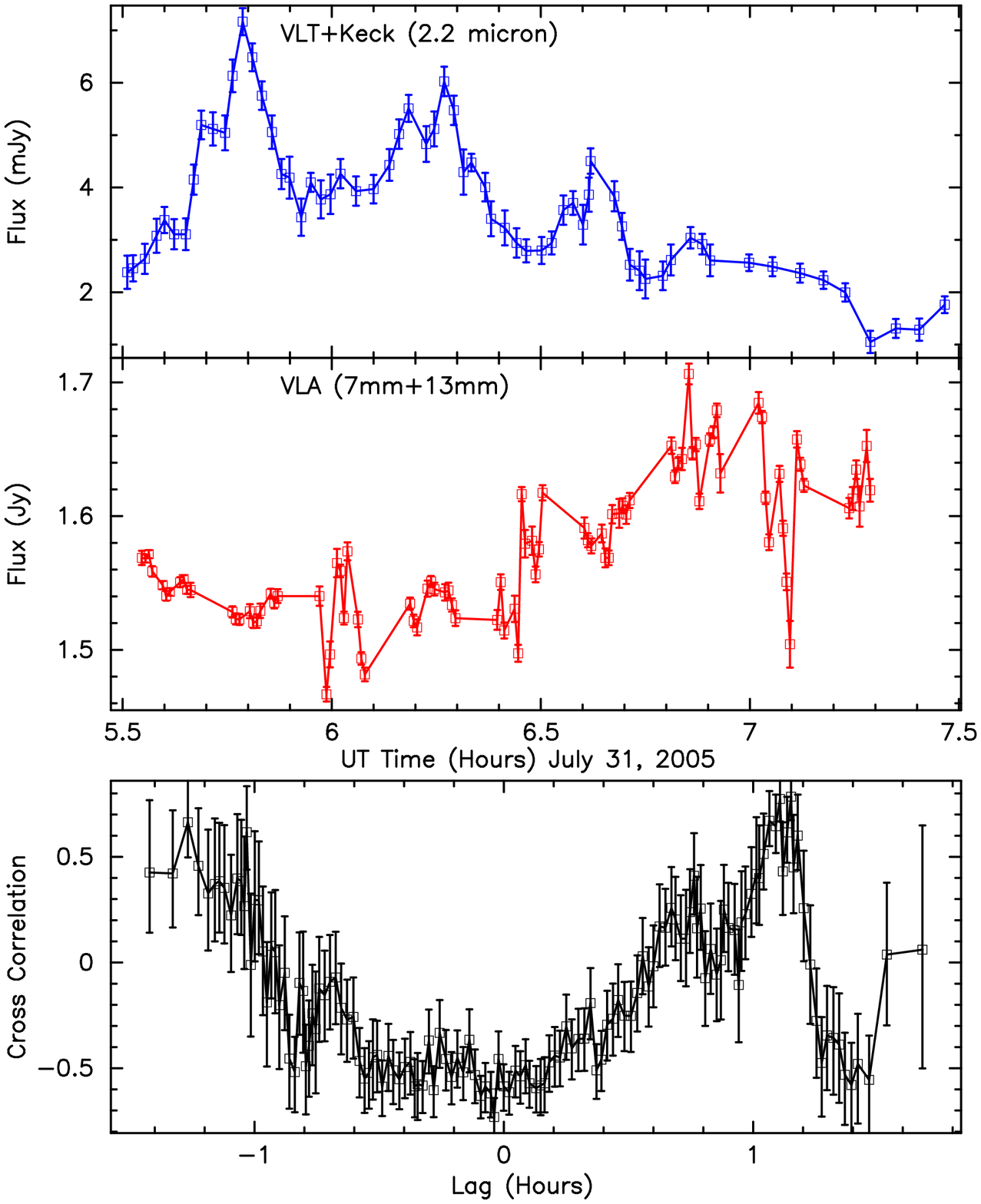}
\caption{
{\it (a) Left} The cross correlation of IR and 7mm light curves is shown
in the bottom panel.  
{\it (b) Right}
Similar to (a) except that radio data has merged the 7 and 13mm data 
by adding  a constant flux of 0.4 Jy to the 13mm data. 
}
\end{figure}

\begin{figure}
\center
\includegraphics[scale=0.4,angle=0]{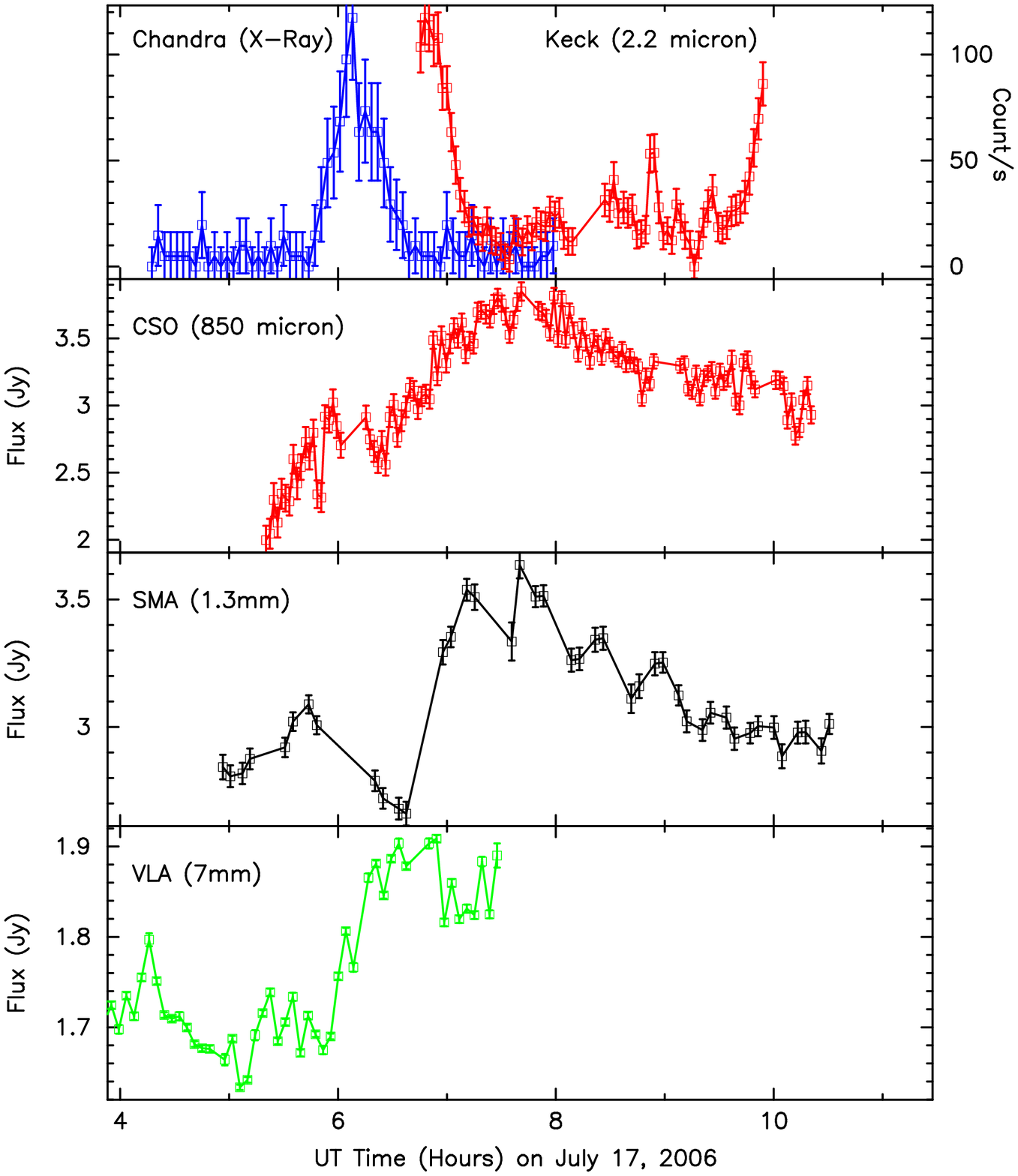}
\includegraphics[scale=0.4,angle=0]{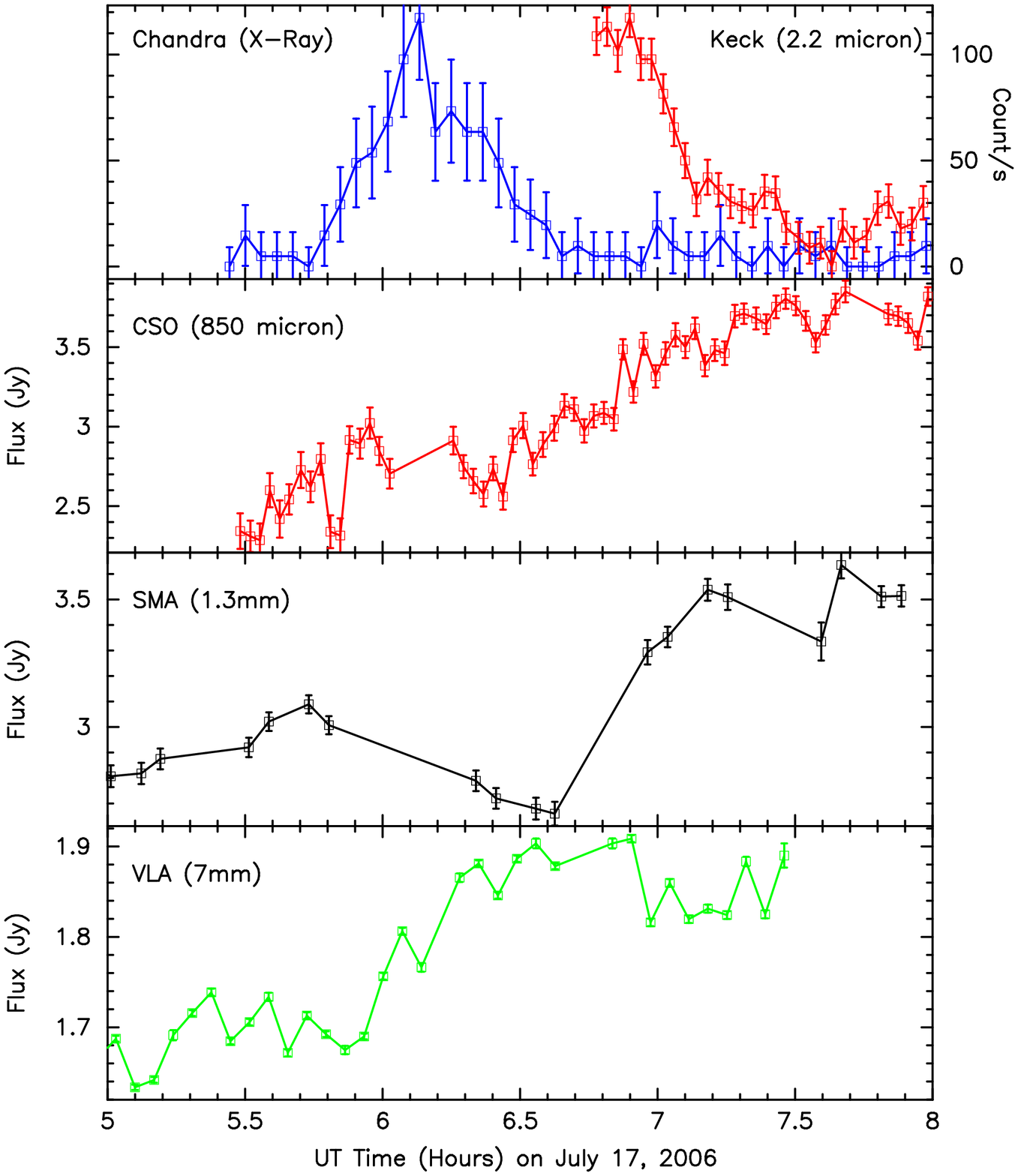}
\caption{
{\it (a) Left }
The light curves of Sgr A*
is shown in five different wavelengths 
The  X-ray (2-8 keV), IR, submm (850 $\mu$m), 1.3mm and  7 mm data are   
binned
every 207sec, 148sec,  120sec, 250sec and  250sec,   respectively 
\citep{hornstein07,marrone08,zadeh08}. 
{\it (b) Right}
The same as (a) except that a close up view of the light curves including a IR data at 
L$'$ band. 
}
\end{figure}

\begin{figure}
\center
\includegraphics[scale=0.4,angle=0]{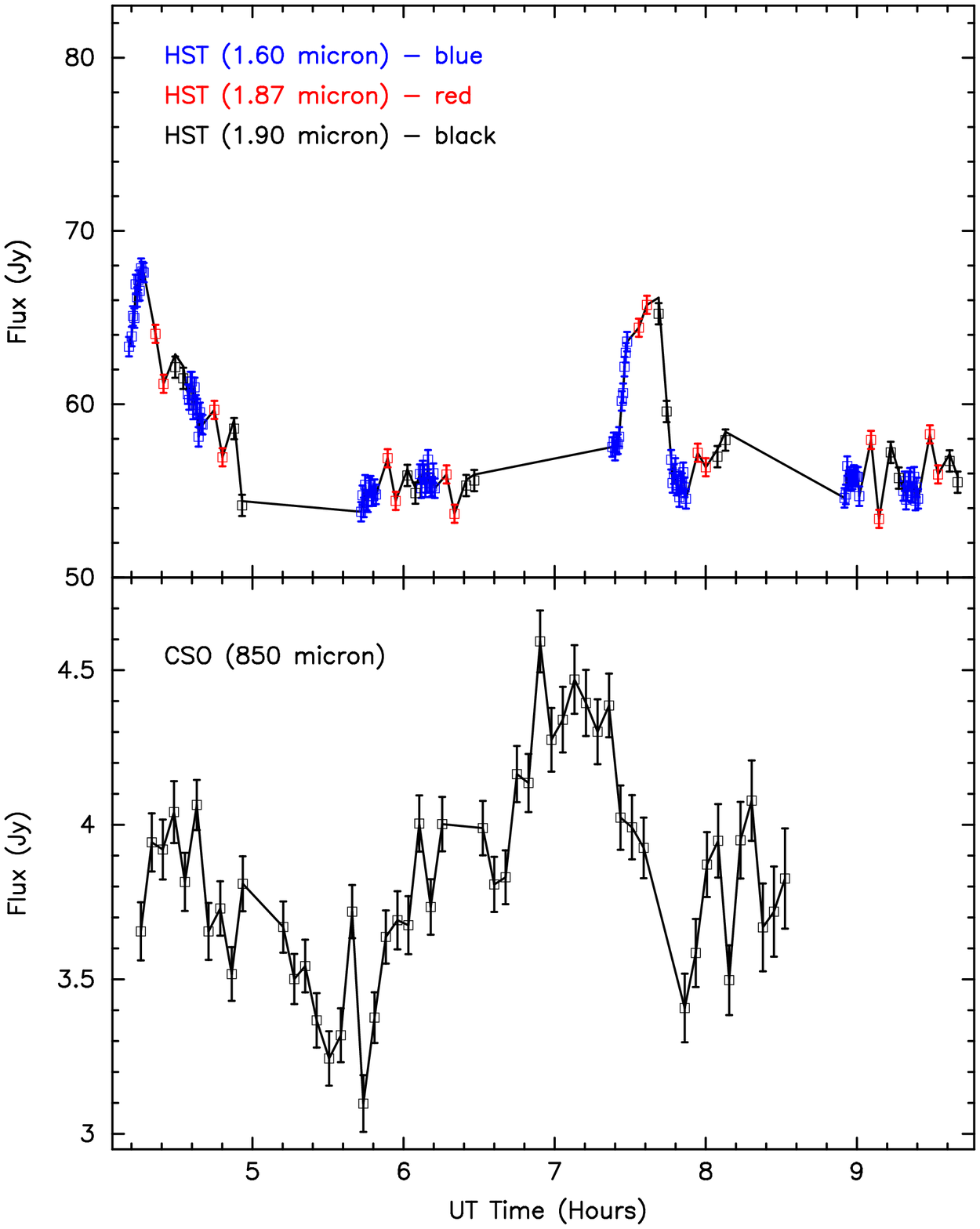}
\includegraphics[scale=0.4,angle=0]{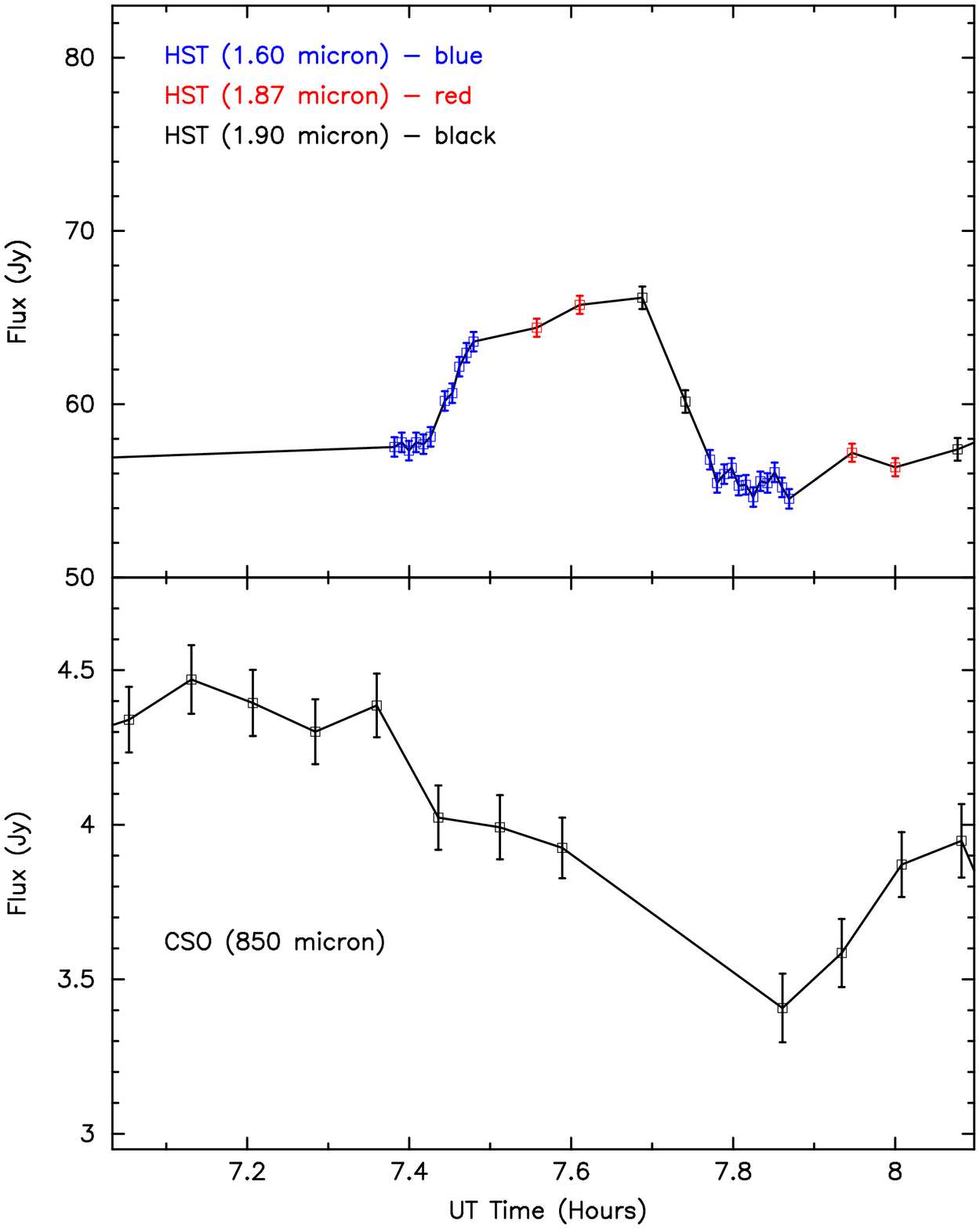}
\caption{
{\it (a) Left}
The simultaneous IR
and 850$\mu$m  light curves using   NICMOS/HST
and CSO  on 2004, September 4.   
The 1.6, 1.87 and 1.90, 850$\mu$m  sampling time is
 32sec, 192sec, 192sec and  280sec, respectively. 
{\it (b) Right}
Similar to (a) except that a close-up view is shown.
}
\end{figure}

\begin{figure}
\center
\includegraphics[scale=1,angle=0]{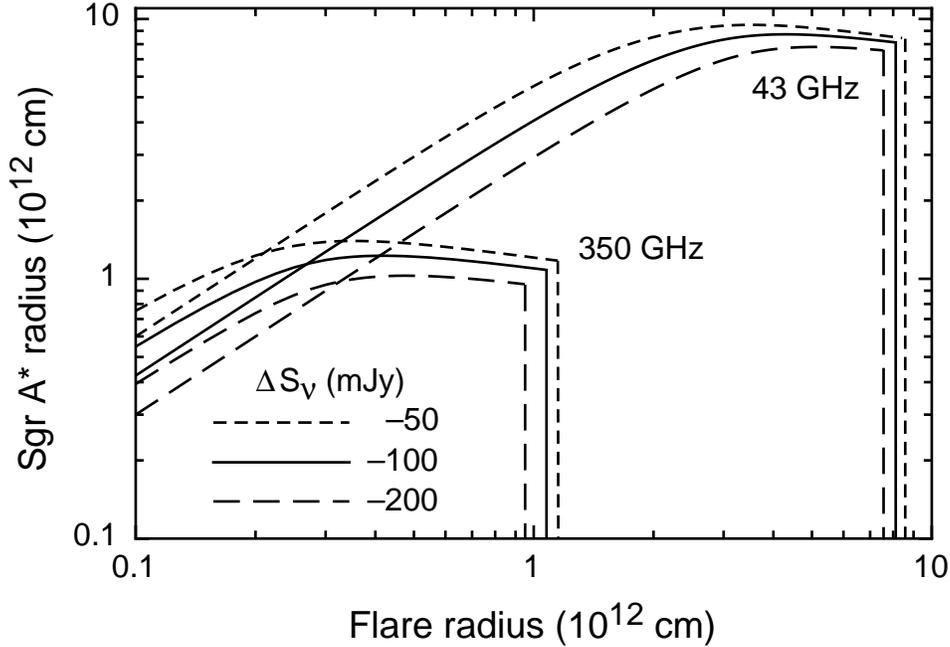}
\caption{Combination of IR flare region radius and Sgr A* quiescent
emission radius needed to produce  drops in flux density at 43 and
350\,GHz of 50, 100 and 200\,mJy (short-dashed, solid, and long-dashed 
curves, respectively.) during an infrared flare.}
\end{figure}

\end{document}